\documentclass[%
 reprint,
 amsmath,amssymb,
 aps,
]{revtex4-1}
\usepackage{graphicx}% Include figure files
\usepackage{dcolumn}% Align table columns on decimal point
\usepackage{bm}% bold math
\begin{document}

\preprint{APS/123-QED}

%\title{\boldmath Enhanced $B_d^0 \to \mu^+\mu^-$ Decay: What if?}

\title{\boldmath Implication of possible observation of
 enhanced $B_d^0 \to \mu^+\mu^-$ decay}

\author{Wei-Shu Hou$^{a}$, Masaya Kohda$^{a}$, and Fanrong Xu$^{b}$}
 \affiliation{$^{a}$Department of Physics, National Taiwan University, Taipei, Taiwan 10617\\
$^{b}$Department of Physics, National Tsing Hua University, Hsinchu, Taiwan 30013}%Lines break automatically or can be forced with \\

%\date{\today}% It is always \today, today,
             %  but any date may be explicitly specified

\begin{abstract}
The very rare $B_d^0 \to \mu^+\mu^-$ decay may be the last chance for
New Physics in flavor sector at the LHC, before the 13 TeV run in 2015.
Partially motivated by the known tension in $\sin2\beta/\phi_1$,
enhancement beyond (3--4)$\times 10^{-10}$ would likely imply
the effect of a fourth generation of quarks. If observed at this level,
the 126 GeV boson may not be the actual Higgs boson, while the $b \to d$
quadrangle (modulo $m_{t'}$) would jump out.
The 2011-2012 data is likely not sensitive to values below 3$\times 10^{-10}$,
and the mode should continue to be pursued with the 13 TeV run.
\begin{description}
%\item[Usage]
% Secondary publications and information retrieval purposes.
\item[PACS numbers]
14.65.Jk % Other quarks (e.g., 4th generations)
12.15.Hh % Determination of Cabibbo-Kobayashi & Maskawa (CKM) matrix elements
11.30.Er % Charge conjugation, parity, time reversal, and other discrete symmetries
13.20.He % Decays of bottom mesons (Leptonic,semileptonic, and radiative decays of mesons)
%\item[Structure]
% You may use the \texttt{description} environment to structure your abstract; use the optional argument of the \verb+\item+ command to give the category of each item.
\end{description}
\end{abstract}

\pacs{Valid PACS appear here}% PACS, the Physics and Astronomy
                             % Classification Scheme.
%\keywords{Suggested keywords}%Use showkeys class option if keyword
                              %display desired
\maketitle

%\tableofcontents

\section{\label{sec:Intro}INTRODUCTION\protect\\}

Despite the discovery~\cite{ATL_H, CMS_H} of a 126 GeV boson by the
Large Hadron Collider (LHC) in 2012, the LHC has so far been a
disappointment: no New Physics beyond the Standard Model (SM) has been seen,
and even the new boson appears Higgs-like, i.e. as prescribed by SM.

Surveying the terrain, there seems one last hope for discovering
New Physics, namely $B_d^0 \to \mu^+\mu^-$.
There is some motivation for enhancement,
from the well known~\cite{Lunghi:2008aa, Buras:2008nn} mild (of order 2$\sigma$)
but lingering tension between direct measurement of CP violation (CPV)
phase of $\bar B_d$--$B_d$ mixing, versus extraction by indirect means.
If an enhanced $B_d^0 \to \mu^+\mu^-$ rate is discovered
with 2011--2012 LHC data, the likely explanation would be
a fourth generation of quarks. This would then cast doubt on
the Higgs boson interpretation of the 126 GeV boson.

The $B_s^0 \to \mu^+\mu^-$ decay has been a highlight pursuit
since Tevatron times,
%with the reach improving by over a factor of 2000~\cite{CDFinal},
and only recently surpassed~\cite{LHCBsmumu} in sensitivity by the LHC.
The drive has been the possibly huge enhancement by exotic scalar effects
inspired by supersymmetry (SUSY), but now excluded by the
first evidence for SM-like rates by the LHCb experiment~\cite{LHCbBsmumu}.
In contrast,
%although sensitivity is better due to larger $B_d$ production rate,
the search for $B_d^0 \to \mu^+\mu^-$ has not shared the limelight.
This is because the SM prediction itself is 30 times lower than
$B_s^0 \to \mu^+\mu^-$. However, the combined LHC bound is now
within~\cite{LHCBsmumu} a factor of 8 of the SM prediction, and
one may ask whether this mode could be anywhere 
enhanced up to this order. 

As pictorialized by the ``Straub plot"~\cite{Straub} and discussed
recently by Stone~\cite{Stone}, most models of enhancement for
$B_d^0 \to \mu^+\mu^-$ have now been eliminated by the SM-like
$B_s^0 \to \mu^+\mu^-$ rate measured by LHCb, with two exceptions.
One is an old, purely left-handed SUSY model~\cite{MSSM-LL}.
However, the region allowed by current data is but a corner of
the parameter space, hence not plausible.
The other would be~\cite{Buras2010} the 4th generation (4G),
where $B_d^0 \to \mu^+\mu^-$ and $B_s^0 \to \mu^+\mu^-$ decays are
modulated by different Cabibbo-Kobayashi-Maskawa (CKM) 
products $V_{t'd}^*V_{t'b}$ and $V_{t's}^*V_{t'b}$,
allowing $B_d^0 \to \mu^+\mu^-$ to be enhanced up to the current bound,
%by an order of magnitude,
even if $B_s^0 \to \mu^+\mu^-$ is SM-like.
Stone has followed conventional wisdom to argue~\cite{Stone}
that 4G has been ``eliminated by the Higgs discovery",
because it ``would cause the Higgs production cross-section to be
nine times larger \ldots"~\cite{4Gover}.
In fact, a comprehensive analysis~\cite{Eberhardt:2012gv} including
electroweak and flavor observables plus earlier Higgs production data
already ruled out 4G in SM framework.
There are two catches in this pessimism, however.
First of all, it is not yet established that the observed 126 GeV
object is the Higgs boson of SM.
For example, a dilaton might mimic~\cite{dilaton} the Higgs with current data.
Second, the Higgs boson of SM does not enter into the $B_d^0 \to \mu^+\mu^-$ process
(the same holds for the $B_d$ box diagram and $B^+ \to \pi^+\mu^+\mu^-$ processes we consider).
To assume indirect arguments in the flavor pursuit is self-defeating,
especially when there is still room for large enhancement;
it actually highlights the potential impact of a discovery.

It was shown~\cite{MHK} recently, through an empirical gap equation~\cite{Gap},
that dynamical electroweak symmetry breaking (DEWSB) could
occur through strong Yukawa coupling of 4G quarks.
Although there is no account for how a dilaton actually emerges, the
scale invariance of this gap equation allows for a dilaton to appear.
The dilaton possibility can be checked experimentally through
the absence, or suppression, of vector boson fusion (VBF) and
associated production (VH) processes,
which requires more data than currently available.
The very large Yukawa coupling needed for DEWSB is consistent with
not finding the 4G quarks so far, where the
current bounds~\cite{4Gbounds} are already
above the nominal~\cite{Chanowitz} unitarity bound (UB). Thus,
the numerical study we present below is only meant as an illustration.

In the following, we review input parameters and constraints,
then present our numerical study.
We indeed find enhancement beyond
$4\times 10^{-10}$ (4 times SM)
is possible~\cite{4times} within the parameter space indicated by the known
tension in $\sin2\Phi_{B_d} \equiv \sin2\phi_1/\beta$.
We give an assessment of immediate and longer term prospects.

\section{Constraints and Input Parameters}

There is no indication for New Physics in $b\to s$ transitions at present.
The best probe is $\sin2\Phi_{B_s}$ measurement pursued by LHCb,
where $\Phi_{B_s}$ is defined as the CPV phase in the $\bar B_s \to B_s$
mixing amplitude (hence $\sin2\Phi_{B_s} \equiv \sin\phi_s$).
This definition is consistent with $\sin2\beta/\phi_1 \equiv \sin2\Phi_{B_d}$
used by the B factories.
The 4G $t'$ quark could have easily affected many $b\to s$ processes~\cite{Buras2010, 4S4G}.
However, all of these, including $s\to d$ transition effects, can be tuned away
or softened by a small $|V_{t's}^*V_{t'b}|$ strength, which is demanded by
$\sin2\Phi_{B_s}$ being consistent with SM expectations and is yet to be measured.
As illustrated by the Straub plot~\cite{Straub, Buras2010}, $B_s\to \mu^+\mu^-$
and $B_d\to \mu^+\mu^-$ can vary independently from each other,
i.e. through $V_{t'd}^*V_{t'b}$ and $V_{t's}^*V_{t'b}$,
subject to constraint from kaon physics (affected by $V_{t'd}^*V_{t's}$).

It is well known~\cite{Lunghi:2008aa, Buras:2008nn},
however, that there is some tension
between the directly measured value~\cite{PDG} of
\begin{equation}
\sin2\beta/\phi_1 = 0.679 \pm 0.020,
\end{equation}
and SM expectation via $\beta/\phi_1 \cong \arg \lambda_t^{\rm SM}$,
where~\cite{simplified}
\begin{equation}
\lambda_t^{\rm SM} = -\lambda_u -\lambda_c
 \simeq -|V_{ud}||V_{ub}|e^{-i\phi_3} + |V_{cd}||V_{cb}|,
\end{equation}
with $\lambda_i \equiv V_{id}^*V_{ib}$. The terms on right-hand side of
Eq.~(2) can be measured at the tree level.
Currently~\cite{PDG},
\begin{equation}
\phi_3 = (68^{+11}_{-10})^\circ,
\end{equation}
and we take the central values $|V_{ud}| = 0.974$,  $|V_{cd}| = 0.23$
and $|V_{cb}| = 0.041$~\cite{PDG}. Variations in these values
are not central to our discussion.

In contrast, $|V_{ub}|$ also has some tension in the measured values.
Extraction via inclusive or exclusive semileptonic
$B$ decays yield approximately $4.41 \times 10^{-3}$ and
$3.23 \times 10^{-3}$~\cite{PDG}, respectively,
with the average value of $4.15 \times 10^{-3}$
(the inclusive approach has better statistics).
We use central values, as our purpose is only for illustration,
hence we will treat the average (which is close to inclusive)
and exclusive cases separately.

Although the strength of $|\lambda_t^{\rm SM}| \simeq 0.0088$
is not sensitive to $|V_{ub}|$, the phase is sensitive to its value,
\begin{align}
\sin2\beta/\phi_1 =
\begin{cases}
0.76  & \text{for}~|V_{ub}|^{\rm ave} \\
0.63  & \text{for}~|V_{ub}|^{\rm excl},
\end{cases}
\end{align}
which both deviate from Eq.~(1) by more than 2$\sigma$
(the inclusive value of 0.81 deviates even more).
This deviation offers some motivation for New Physics
in $b \to d$ transitions.
It could easily be due to the 4G quark $t'$, where one simply
augments Eq.~(2) by
\begin{equation}
\lambda_t = \lambda_t^{\rm SM} -\lambda_{t'},
\end{equation}
and the $b\to d$ triangle becomes a quadrangle
\begin{equation}
\lambda_u + \lambda_c + \lambda_t + \lambda_{t'} = 0.
\end{equation}
In our following study, we parameterize~\cite{Hou:2005hd}
\begin{equation}
\lambda_{t'} = r_{db}\,e^{i\phi_{db}}.
\end{equation}
In our phase convention,
$\lambda_c = V_{cd}^*V_{cb}$ is practically real,
while $\lambda_u = V_{ud}^*V_{ub}$ is basically
the same as in SM.

To study $\sin2\Phi_{B_d}$ and ${\cal B}(B_d \to \mu^+\mu^-)$
in the $r_{db}$--$\phi_{db}$ plane, other constraints should be considered:
\begin{itemize}
\item radiative $b\to d\gamma$ processes (including $B\to \rho\gamma$)
is ineffective because it is hard to separate from $b\to s\gamma$,
difficult to study with LHCb, and in any case insensitive to virtual 4G
effects;
\item $B \to \pi\pi$ decays, while quite well studied, suffers from hadronic
effects (even $B\to K\pi$ suffers from hadronic effects),
and do not provide good constraints;
\item the well measured $\Delta m_{B_d}$ provides a constraint
    through uncertainties in $f_{B_d}^2\hat B_{B_d}$;
\item only very recently was the electroweak penguin $B^+\to \pi^+\mu^+\mu^-$ decay measured~\cite{LHCb:2012de}, in contrast to electroweak $b\to s$ penguins.
\end{itemize}
Although it may be a little surprising, there are not many
observables that provide sound constraints on $\lambda_{t'}$.
We collect below the relevant formulas for our study.

The $t'$ effect to $B_d$ mixing
\begin{align}
&\Delta m_{B_d} %\simeq 2|M_{12}^d|
 \simeq \frac{G_F^2M_W^2}{6\pi^2}m_{B_d}\hat B_{B_d}f_{B_d}^2 \eta_B |\Delta_{12}^d|, \nonumber \\
&\sin 2\Phi_{B_d} %\equiv \sin (\arg M_{12}^d)
 \simeq \sin (\arg \Delta_{12}^d),
\end{align}
is (explicit forms can be found in Ref.~\cite{Hou:2006mx})
\begin{align}
\Delta_{12}^d \equiv&\ (\lambda_t^{\rm SM})^2 S_0(x_t) \nonumber \\
  &+ 2\lambda_t^{\rm SM}\lambda_{t^\prime} \Delta S_0^{(1)}
  + \lambda_{t^\prime}^2 \Delta S_0^{(2)}, \\
\Delta S_0^{(1)} \equiv&\ \tilde S_0(x_t,x_{t^\prime}) - S_0(x_t),\\
\Delta S_0^{(2)} \equiv&\ S_0(x_{t^\prime}) -2 \tilde S_0(x_t,x_{t^\prime}) +S_0(x_t),
\end{align}
where $x_i = m_i^2/M_W^2$.
Besides 4G parameters, the main uncertainty is in~\cite{Laiho:2009eu}
\begin{equation}
f_{B_d}\hat B_{B_d}^{1/2} = (227 \pm 19)\ {\rm MeV}.
\end{equation}

\begin{figure*}[t!]
\centering
%\includegraphics[width=60mm,height=40mm]{plots/twidth.pdf}
%\vspace{30mm}
{\includegraphics[width=67mm]{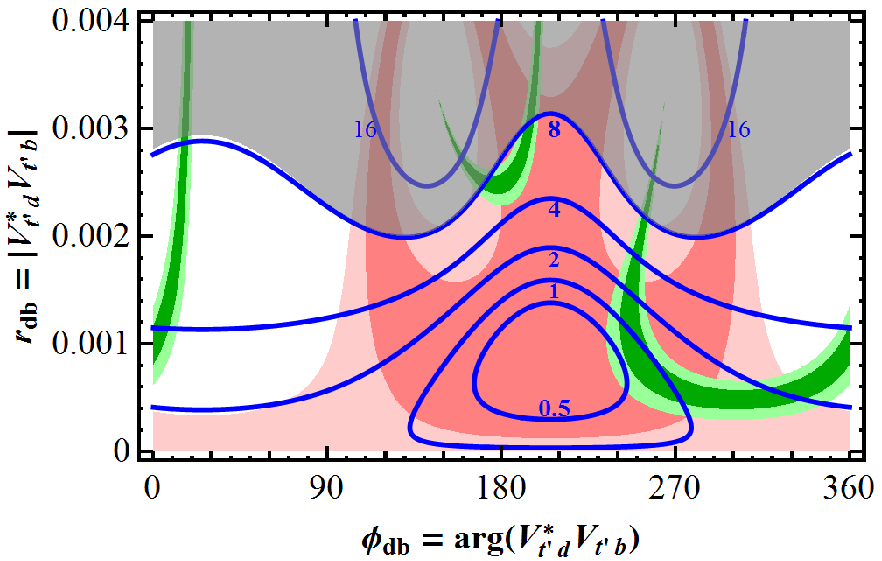}
 \includegraphics[width=67mm]{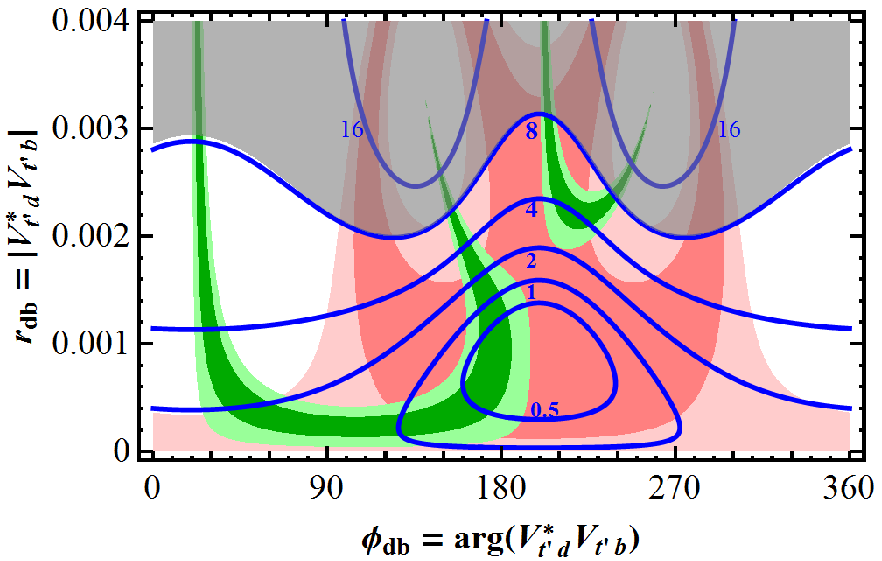}
}
\caption{
Allowed region in $|V_{t'd}^*V_{t'b}|$--$\arg V_{t'd}^*V_{t'b}$
(i.e. $r_{db}$--$\phi_{db}$) plane for
(a) average (b) exclusive $|V_{ub}|$ values, for $m_{t'}= 700$ GeV.
The solid-blue lines are labeled $10^{10}{\cal B}(B_d \to \mu^+\mu^-)$ contours,
where above the value of 8 (semi-transparent gray) is excluded by the combined result
of LHC experiments.
The dark (light) narrow green-shaded contours
 correspond to the 1(2)$\sigma$ regions of $\sin2\Phi_{B_d}$ (Eq.~(1)),
while the broad pink-shaded contours
 correspond to the 1(2)$\sigma$ regions of $\Delta m_{B_d}$ allowed by Eq.~(12).
} \label{DeltamBd-700}
\end{figure*}

\begin{figure*}[t!]
\centering
%\includegraphics[width=60mm,height=40mm]{plots/twidth.pdf}
%\vspace{30mm}
{\includegraphics[width=67mm]{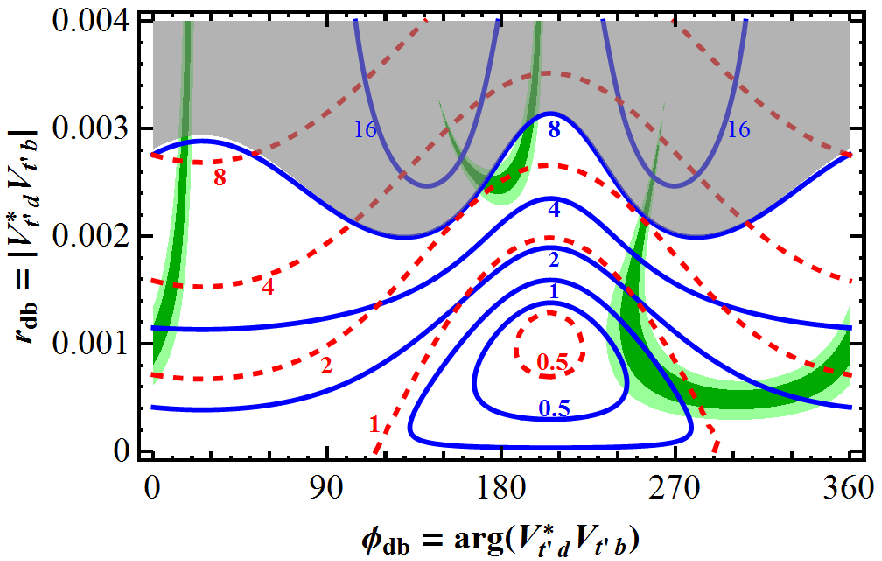}
 \includegraphics[width=67mm]{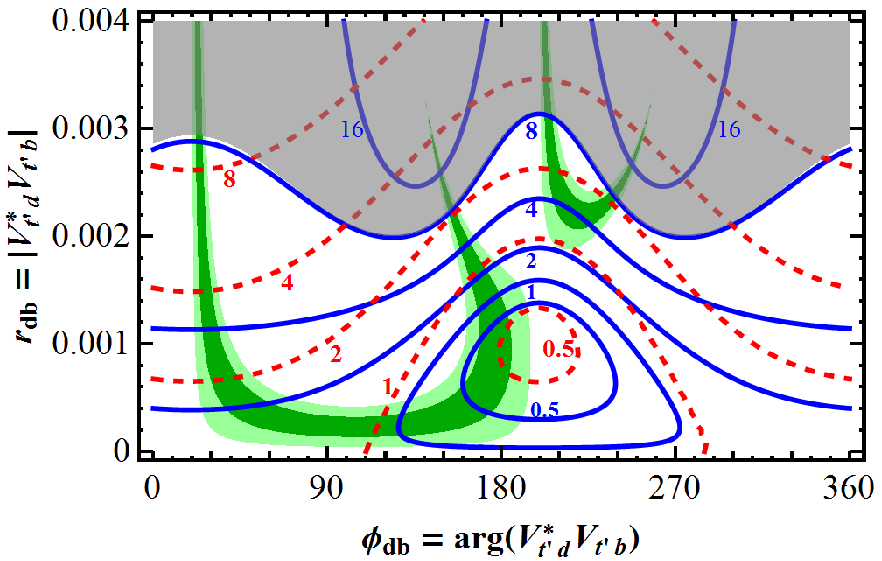}
}
\caption{
Same as Fig.~1, but with $\Delta m_{B_d}$ allowed regions
replaced by the contours (red-dashed) of ratio of 4G over SM branching ratios
for $B^+\to \pi^+\mu^+\mu^-$, integrated over the $q^2$ range of
1--6 GeV$^2$.
} \label{B2pimumu-700}
\end{figure*}

For the current bound~\cite{LHCBsmumu} of
\begin{equation}
{\cal B}(B_d \to \mu^+\mu^-) < 8.1\times 10^{-10},
\end{equation}
our purpose is to illustrate whether, and how, it could
get enhanced to such values by 4G effect.
Here, we use the usual trick \cite{Buras:2003td} of 
``normalizing" the
branching ratio,
\begin{align}
&\hat{\mathcal B}(B_d \to \mu^+\mu^-)
\equiv \frac{\mathcal B (B_d \to \mu^+\mu^-)}{\Delta m_{B_d}}
 \Delta m_{B_d}^{\rm exp} \nonumber \\
&= C \frac{\tau_{B_d}\Delta m_{B_d}^{\rm exp}}{\hat B_{B_d}}\frac{\eta_Y^2}{\eta_B}
 \frac{ \left| \lambda_t^{\rm SM} Y_0(x_t) +\lambda_{t^\prime}\Delta Y_0 \right|^2}
{| \Delta_{12}^d |}
\end{align}
where $\Delta Y_0 = Y_0(x_{t'}) - Y_0(x_t)$ with
$Y_0(x)$ given in Ref.~\cite{Buras2010}, and
\begin{align}
C
= 6\pi \left(\frac{\alpha}{4\pi \sin^2\theta_W}\right)^2
 \frac{m_\mu^2}{M_W^2}.
\end{align}
Through the ratio of Eq.~(14), one not only eliminates the
hadronic parameter $f_{B_d}$, but the $\lambda_t^{\rm SM}$ factor
also cancels in the SM case, and one recovers the SM
result of $1.1 \times 10^{-10}$, with little sensitivity to $|V_{ub}|$.

The treatment of $B^+\to \pi^+\mu^+\mu^-$ would be given
in the next section.

\section{Phenomenological Study with Heavy $t'$}

We plot in Fig.~1 for $m_{t'} = 700$ GeV the 2$\sigma$ range
in the $r_{db}$--$\phi_{db}$ plane, for
$\sin2\Phi_{B_d}$ (green) allowed~\cite{cos} by experimental measurement of Eq.~(1),
$\Delta m_{B_d}$ (pink) allowed by lattice error in Eq.~(12), and
the bound on $B_d \to \mu^+\mu^-$ (gray exclusion) according to Eq.~(13).
We include labeled contours of 0.5, 1, 2, 4, 8 for $10^{10}{\cal B}(B_d \to \mu^+\mu^-)$.
Fig.~1(a) and (b) are for taking $|V_{ub}|$ to be the central values of
$4.15 \times 10^{-3}$ and $3.23 \times 10^{-3}$, respectively, for
the mean (between inclusive and exclusive) and exclusive values
from semileptonic $B$ decay studies.

Consider Fig.~1(a), i.e. for $|V_{ub}| = 4.15\times 10^{-3}$,
the average between inclusive and exclusive measurements
(the inclusive case is qualitatively similar).
The well measured CP phase $\sin2\Phi_{B_d}$ is
sensitive to $t'$ effects, but free from hadronic uncertainties,
hence the narrow (green) contour bands. In contrast,
$\Delta m_{B_d}$ is less sensitive to $\phi_{db}$,
and more accommodating because of
hadronic uncertainty in $f_{B_d}\hat B_{B_d}^{1/2}$.
The broad (pink) contour bands show the 1 and 2$\sigma$ allowed region by Eq.~(12),
and rules out a branch of the $\sin2\Phi_{B_d}$ contour
(for $\phi_{db}$ between $-10^\circ$ to $15^\circ$), due to
coherent enhancement of $\Delta m_{B_d}$ from $t'$ effects.

\begin{figure*}[t!]
\centering
%\includegraphics[width=60mm,height=40mm]{plots/twidth.pdf}
%\vspace{30mm}
{\includegraphics[width=67mm]{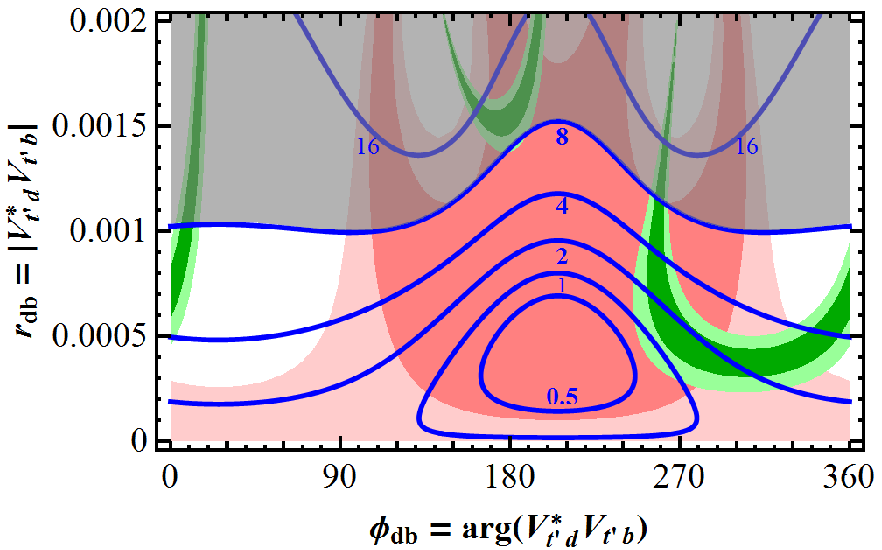}
 \includegraphics[width=67mm]{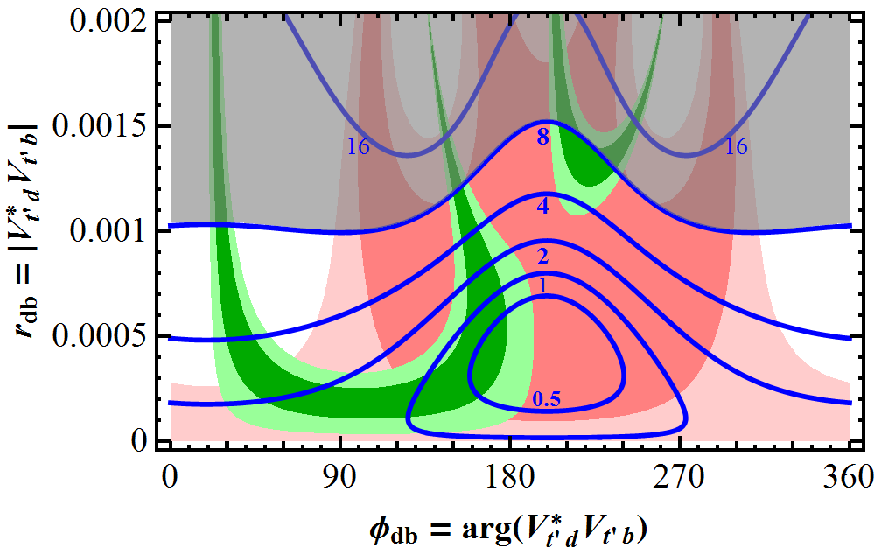}
}
\caption{
Same as Fig.~1, but for $m_{t'}= 1000$ GeV.
} \label{DeltamBd-1000}
\end{figure*}

\begin{figure*}[t!]
\centering
%\includegraphics[width=60mm,height=40mm]{plots/twidth.pdf}
%\vspace{30mm}
{\includegraphics[width=67mm]{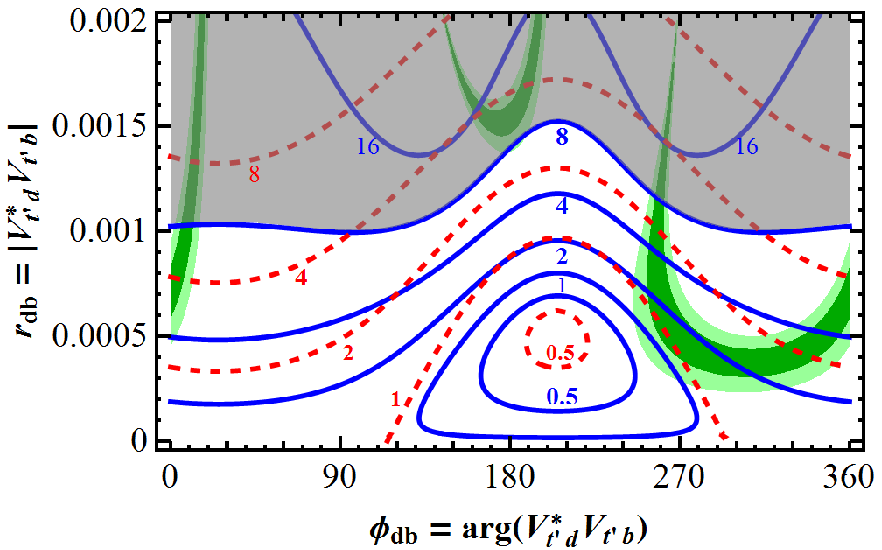}
 \includegraphics[width=67mm]{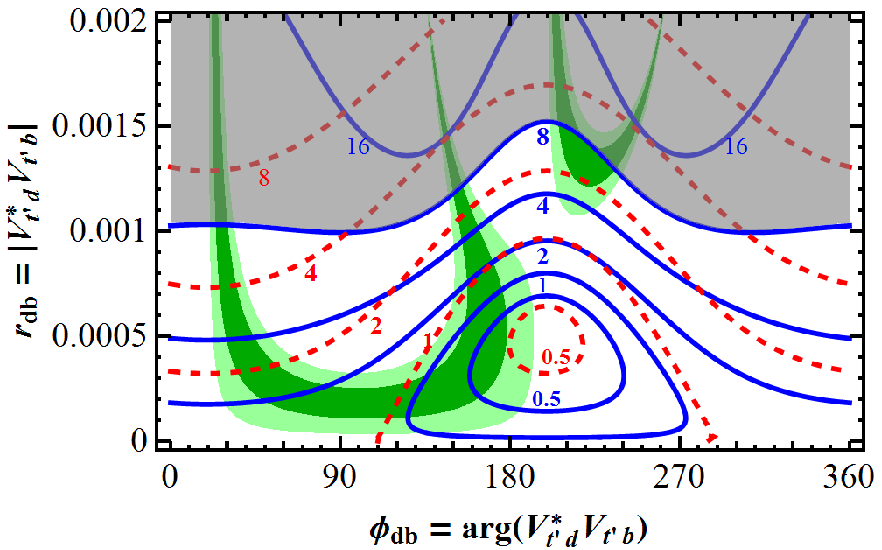}
}
\caption{
Same as Fig.~2, but for $m_{t'}= 1000$ GeV.
} \label{B2pimumu-1000}
\end{figure*}

Consider now the gray excluded region from the combined LHC
bound on $B_d \to\mu^+\mu^-$, Eq.~(13). It is seen that
there are two slivers of parameter space, around
$(r_{db},\ \phi_{db}) \sim (0.0025,\ 180^\circ)$ (region A)
and $(0.002,\ 252^\circ)$ (region B),
where ${\cal B}(B_d \to\mu^+\mu^-)$ could be above $4 \times 10^{-10}$,
or enhanced by 4 times over SM,
which are discovery zones for 2011-2012 LHC data.
Near region B, ${\cal B}(B_d \to\mu^+\mu^-)$ quickly drops
below $4 \times 10^{-10}$ as $r_{db}$ becomes weaker than 0.002.
For $\phi_{db} \sim 245^\circ$ and $r_{db}$ varying from 0.0008 to 0.0015,
${\cal B}(B_d \to\mu^+\mu^-)$ hovers at (1--2)$\times 10^{-10}$,
while for $r_{db} \sim 0.0004$ to 0.0008 and $\phi_{db}$ varying from
$240^\circ$ to $330^\circ$, ${\cal B}(B_d \to\mu^+\mu^-)$ hovers at
(0.5--2)$\times 10^{-10}$, i.e. within a factor of two of SM expectations.
These regions, combining to a broad crescent shape which we refer to as
``region C", would likely need much more data to probe.

The LHCb experiment has recently measured~\cite{LHCb:2012de}
\begin{align}
{\cal B}(B^+\to \pi^+\mu^+\mu^-)
= (2.3\pm 0.6 \pm 0.1) \times 10^{-8},
\end{align}
which is the rarest $B$ decay observed to date.
The result is consistent with SM expectations, but interpretation
depends on form factor models.
To reduce form factor dependence, we take the ratio
\begin{align}
R_{\pi\mu\mu} \equiv
\frac{\mathcal B(B^+\to \pi^+\mu^+\mu^-)|_{\rm 4G}}
{\mathcal B(B^+\to \pi^+\mu^+\mu^-)|_{\rm SM}},
\end{align}
where both 4G and SM results are integrated from $q^2 = (1,\ 6)\ {\rm GeV}^2$,
which is under better numerical control~\cite{Beneke:2001at, Beneke:2004dp}.
Since this does not match what LHCb does, we draw contours
in Fig.~2 (red-dashed), and view $R_{\pi\mu\mu} \sim 2$--3 as the range
beyond which LHCb would have found inconsistency with SM expectations.
Thus, we are interpreting LHCb's statement of
consistency with SM, allowing for form factor uncertainties.
It is clear that this approach is not as good as
the zero crossing point $q_0^2$ for $A_{\rm FB}(B\to K^*\mu\mu)$,
but this is the first observation of rare $b\to d\ell\ell$ decays,
compared to the decade-long exploration of $b\to s\ell\ell$ processes.
For numerics, we combine Wilson coefficients at next-to-leading order
with leading order decay amplitude based on the QCD factorization %(QCDF)
approach \cite{Beneke:2001at, Beneke:2004dp}.
For dealing with New Physics, and as we take a ratio,
this should suffice for our purpose.

If we now compared Fig.~1(a) with Fig.~2(a), we see that
$\Delta m_{B_d}$ is more powerful than ${\cal B}(B^+\to \pi^+\mu^+\mu^-)$
in excluding the $\sin2\Phi_{B_d}$-allowed branch near $\phi_{db} \sim 0$.
This is reasonable, since $B^+\to \pi^+\mu^+\mu^-$ is only recently
observed and prone to hadronic form factor uncertainties, while
$\Delta m_{B_d}$ has been measured since 25 years, with hadronic
uncertainty narrowed down to $f_{B_d}\hat B_{B_d}^{1/2}$, which
itself has been subject to intense lattice studies for years.
It is, however, comforting to see that for region A, $R_{\pi\mu\mu}$ is
not more than 2 (except the upper reach near $\phi_{db} \sim 190^\circ$),
hence should be easy to accommodate by form factors, while for regions B
and especially region C, $R_{\pi\mu\mu}$ is even less than 2 and closer to 1.
Thus, the newly measured $B^+\to \pi^+\mu^+\mu^-$ does provide a sanity check.

Turning to the case of exclusive $|V_{ub}|$ value, Fig.~1(b) and 2(b),
we find that regions A and B basically switch roles. This is because
for $|V_{ub}| \sim 3.23 \times 10^{-3}$, the expected $\sin2\Phi_{B_d}$
value in SM falls below that of direct measurement, as seen in
comparing Eq.~(4) to Eq.~(1).
Calling it region A$'$, the sliver of region around
$(r_{db},\ \phi_{db}) \sim (0.002,\ 160^\circ)$ could enhance
${\cal B}(B_d \to\mu^+\mu^-)$ more than 4 times above SM, and
observable with present LHC data. Region A$'$ extends to
the broad crescent region C$'$, where even $r_{db}$ values
as lower as 0.0002 could account for the measured $\sin2\Phi_{B_d}$,
but ${\cal B}(B_d \to\mu^+\mu^-)$ can be probed only beyond 2015.
Again, $\Delta m_{B_d}$ excludes the $\sin2\Phi_{B_d}$-allowed
branch around $\phi_{db}\sim 30^\circ$.
Region B$'$ is now a considerably broader region in parameter space
that allows enhancement of ${\cal B}(B_d \to\mu^+\mu^-)$
above $4\times 10^{-10}$. For example,
for $r_{db}$ above 0.0023 and $\phi_{db}$ above 230$^\circ$,
${\cal B}(B_d \to\mu^+\mu^-)$ can be greater than $6\times 10^{-10}$,
$f_{B_d}\hat B_{B_d}^{1/2}$ is within 2$\sigma$ of Eq.~(12),
while $R_{\pi\mu\mu}$ is not more than 2.
We also see that, for region B$'$, $R_{\pi\mu\mu}$ provides as good,
perhaps better constraint, than $\Delta m_{B_d}$, disfavoring
the region of $r_{db}$ greater than 0.0025 around $\phi_{db}\sim 205^\circ$,
that seems perfectly allowed by $\Delta m_{B_d}$.

\begin{figure*}[t!]
\centering
%\includegraphics[width=60mm,height=40mm]{plots/twidth.pdf}
%\vspace{30mm}
{\includegraphics[width=130mm]{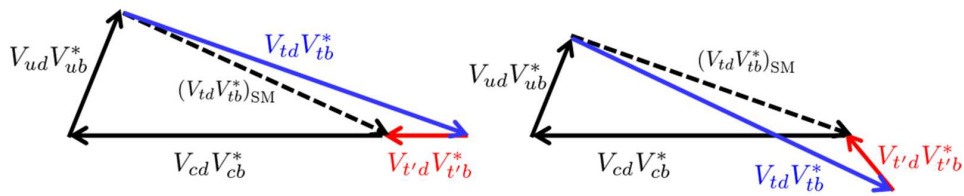}
}
\caption{
Sample $b\to d$ quadrangles
for $\lambda_{t^\prime}=V_{t^\prime d}^*V_{t^\prime b}=0.0025\, e^{i 180^\circ}$
with average $|V_{ub}|=4.15\times 10^{-3}$ (left), and
for $\lambda_{t^\prime}=V_{t^\prime d}^*V_{t^\prime b}=0.0023\, e^{i 230^\circ}$
with exclusive $|V_{ub}|=3.23\times 10^{-3}$ (right).
} \label{fig:BdQuad}
\end{figure*}

Now let us consider $m_{t'}$ values.
The 700 GeV value used so far is just above current experimental limits~\cite{4Gbounds},
and correspond to Yukawa coupling strength $y_{t'} \simeq 4$,
or $\alpha_{t'} \simeq 1.3$, which is why there is UB violation (UBV).
However, we do not quite know what is the true expansion parameter.
Furthermore, even if perturbation breaks down, it does not mean there is no $t'$ effect.
In fact, perturbation in $\lambda_{t'}$ certainly holds,
though the functions $\Delta S_0^{(i)}$ and $\Delta Y_0$ in
Eqs.~(9) and (14) gets modified by UBV effects. %but
The overall form of these equations should not change.
We therefore consider the $m_{t'} = 1000$ GeV case,
i.e. $\alpha_{t'} \simeq 2.6$,
to illustrate the situation far beyond UBV~\cite{Chanowitz}.
Note that Ref.~\cite{MHK} finds DEWSB occurs for $y_{Q}$ %$\lambda_{Q}$
(the 4G doublet is treated as very close to degenerate)
of order $4\pi$, i.e. of order the $\pi NN$ coupling,
implying 4G quark masses no less than 2 TeV!

The plots corresponding to Figs.~1 and 2, but with
$m_{t'} = 1000$ GeV, are given in Figs.~3 and 4.
We generally see reduced $r_{db}$ values.
Region A is now excluded, but regions B, A$'$ and B$'$ become
more robust in $\Delta m_{B_d}$, but values for
${\cal B}(B_d \to\mu^+\mu^-)$ higher than (5--6)$\times 10^{-10}$
are slightly disfavored by $B^+\to \pi^+\mu^+\mu^-$.
Viewed differently, if enhanced $B_d \to\mu^+\mu^-$ is discovered,
one may try to scrutinize whether $B^+\to \pi^+\mu^+\mu^-$ is
also somewhat enhanced beyond SM.
Regions C and C$'$ generally stand well,
with at best mildly enhanced $B_d \to\mu^+\mu^-$.

\section{\label{sec:Discussion} \boldmath
Discussion and Conclusion\protect\\}

We started with the question of what could still enhance
$B_d \to\mu^+\mu^-$ decay, when everything at the LHC seems consistent with SM.
The answer is that, probably only the 4G $t'$ quark could do the job, even if
4G seems disfavored by the Higgs-like nature of the 126 GeV boson.
Admittedly, even if 4G is the explanation for the $\sin2\Phi_{B_d}$
tension as seen by the B factories, to have ${\cal B}(B_d \to\mu^+\mu^-)$
to be within a factor of 2 of the current bound of $8.1\times 10^{-10}$
is only a fraction of the allowed parameter space,
hence not particularly likely.
However, only with such enhancement is there any chance for LHC experiments
to make the discovery with 2011-2012 data, and discovery it indeed will be.
If discovered --- within 2013 --- then not only 4G would get uplifted,
some doubt would be cast on the SM Higgs nature of the 126 GeV boson,
while ``impostors" such as dilaton would gain in weight. We have
remarked in the Introduction that it would take the establishment of
VBF and VH production processes to exclude the dilaton possibility,
which cannot be achieved with 2011-2012 data~\cite{MHK}.

An intriguing outcome of discovering $B_d \to\mu^+\mu^-$ decay
would be that, all of a sudden, the $b\to d$ triangle falls into our lap!
Let us illustrate.
Since $m_{t'} = 1000$ GeV cases have smaller
$r_{db} \equiv |\lambda_{t'}| \equiv |V_{t'd}^*V_{t'b}|$ values,
for reasons of plotting, we take two examples from $m_{t'} = 700$ GeV.
From region A of Fig.~1(a) (average $|V_{ub}| = 4.15 \times 10^{-3}$),
we take $\lambda_{t'} = V_{t'd}^*V_{t'b} = 0.0025\, e^{i 180^\circ}$.
From region B of Fig.~1(b) (exclusive $|V_{ub}| = 3.23 \times 10^{-3}$),
we take $\lambda_{t'} = V_{t'd}^*V_{t'b} = 0.0023\, e^{i 230^\circ}$.

The quadrangle of Eq.~(6) is constructed as follows.
To simplify discussions, we normalize to $\lambda_c = V_{cd}^*V_{cb} = -0.0094$,
which becomes a unit vector pointing left.
Then, $\hat \lambda_u = V_{ud}^*V_{ub}/|\lambda_c| = 0.44\,e^{-i 68^\circ}$,
$0.34\,e^{-i 68^\circ}$, respectively, for the average and inclusive cases,
with corresponding $\hat \lambda_{t'} = 0.27\, e^{i 180^\circ}$,
$0.24\, e^{i 230^\circ}$. Then $\hat \lambda_t$ just connects the
tip of $\hat \lambda_u$ with the end of $\hat \lambda_{t'}$.
The two examples for 700 GeV are plotted in Fig.~\ref{fig:BdQuad}
in the form to compare with the usual SM triangle~\cite{PDG}.
These are relatively precise quadrangles, and illustrate how
4G accounts for a shift in $\sin2\Phi_{B_d}$ away from SM expectation,
where $\Phi_{B_d}^{\rm SM}$ is the angle between the dashed line,
$\lambda_t^{\rm SM}$ and the real axis.
Since $t'$ is much heavier than $t$, a smaller $\lambda_{t'}$
could cause the shift.

The sample $b\to d$ quadrangles are for largest allowed solutions
for $r_{db}$, i.e. regions A (for $|V_{ub}|^{\rm ave}$)
and B$'$ (for $|V_{ub}|^{\rm excl}$) for $m_{t'} = 700$ GeV,
and would be the case if $B_d \to\mu^+\mu^-$ is discovered soon.
They are relatively extreme, however, since even for $m_{t'} = 700$ GeV,
regions C and C$'$ can provide solutions for $\sin2\Phi_{B_d}$
for much smaller $r_{db} = |V_{t^\prime d}^*V_{t^\prime b}|$ values,
with possible phase values extending over a large range.
For heavier $t'$ illustrated by 1000 GeV,
$|V_{t^\prime d}^*V_{t^\prime b}|$ is smaller by half compared
to 700 GeV case, with region A is eliminated.

The quadrangles of Fig.~\ref{fig:BdQuad}
reminds us of the possible~\cite{Hou:2008xd}
link to the baryon asymmetry of the Universe (BAU):
4G greatly enhances CPV from SM, and is seemingly sufficient for BAU
(although a first order phase transition remains an issue),
which boosts the merit of 4G. It does not depend much on the area of
the quadrangle, as the enhancement rests in powers of $m_{t'}$ and $m_{b'}$.
We note that $\lambda_{t'}$ in Fig.~\ref{fig:BdQuad}, though smaller in strength
than $\lambda_t$ and $\lambda_c$, is not that small compared with $\lambda_u$.
Furthermore, we know that $|V_{t'b}|$ cannot be more than 0.1~\cite{Hou:2010mm},
especially for our large $m_{t'}$ values.
Hence, $|\lambda_{t'}|$ plotted in Fig.~\ref{fig:BdQuad}
correspond to $|V_{t'd}|$ that is larger than $|V_{td}|^{\rm SM} \simeq 0.0088$,
which does not fit the CKM pattern of trickling off as one goes further off-diagonal.
One could use this to argue that enhanced $B_d \to\mu^+\mu^-$ decay
to the level observable with 2011-2012 data is not plausible.
However, the issue is best left to experiment.

For $m_{t'} = 1000$ GeV, $|\lambda_{t'}|$ values tend to drop by half,
but $|V_{t'd}|$ would still be comparable to $|V_{td}|$.
Only if one gives up enhancement would the ratio $|V_{t'd}/V_{td}|$ turn ``natural".
In fact, for the exclusive value case for $V_{ub}$,
$|\lambda_{t'}|$ (i.e. $r_{db}$) could be (1--2)$\times 10^{-4}$
and still account for $\sin2\Phi_{B_d}$ ``anomaly".
Such values for $|V_{t'd}|$ would become ``natural" when compared with $|V_{td}|$.
%~\cite{natural-btos}.
However, even if 4G gains support by 2015,
this region (C and C$'$) would need a very large data set to explore.

We conclude that 2013 remains a pivotal year where
one could discover the very rare $B_d \to\mu^+\mu^-$ decay mode
at over 4 times SM expectations. The chance is not large,
but not zero either,
with partial motivation from the (mild) $\sin2\Phi_{B_d}$ discrepancy.
If discovered with 2011-2012 data set, the implications would be quite huge:
uplifting the 4th generation (with prospect of CPV for BAU),
casting some doubt on the SM Higgs interpretation of the 126 GeV boson,
and perhaps the only New Physics (at least in flavor sector)
uncovered at the 7 and 8 TeV runs at the LHC.
But it is more likely that the LHC would once again push the limits down towards SM.
If such is the case, the fate of the 4G would have to be determined elsewhere.
But $B_d \to\mu^+\mu^-$ should certainly be pursued further at the 13 TeV run.

\vskip0.3cm
\noindent{\bf Acknowledgement}.  WSH is supported by the the
Academic Summit grant NSC 101-2745-M-002-001-ASP of the
National Science Council, as well as by grant NTU-EPR-102R8915.
MK is supported under NTU-ERP-102R7701 and the Laurel program,
and FX under NSC 101-2811-M-007-051.


\begin{thebibliography}{9}
\bibitem{simplified}
  Our discussion is rather simplified compared with making
  global fits to all data other than the direct measurement of
  $\sin2\phi_1/\beta$. This is in part because our purpose
  is only for illustration. The other reason is because
  $\phi_3$ itself, like $|V_{ub}|$ is now becoming directly measured.
\bibitem{cos}
  We have used $\cos2\Phi_{B_d} > 0$~\cite{PDG}
  to eliminate some of the solution branches.
\end{thebibliography}
\end{document}